\documentclass[journal]{IEEEtran}
\usepackage{amsmath,amsfonts}
\usepackage{algorithmic}
\usepackage{array}
\usepackage[caption=false,font=normalsize,labelfont=sf,textfont=sf]{subfig}
\usepackage{textcomp}
\usepackage{stfloats}
\usepackage{url}
\usepackage{bbm}
\usepackage{verbatim}
\usepackage[normalem]{ulem}
\usepackage{xcolor}
\usepackage{booktabs}
\usepackage{graphicx}
\usepackage{multirow}
\usepackage{amssymb}
\usepackage{acro}
\usepackage{bbding}
\usepackage{diagbox}
\usepackage{balance}
\usepackage{cite}
\usepackage{arydshln}
\usepackage{silence}

\def\BibTeX{{\rm B\kern-.05em{\sc i\kern-.025em b}\kern-.08em
    T\kern-.1667em\lower.7ex\hbox{E}\kern-.125emX}}

\author{
\IEEEauthorblockN{Ruijie Tao, Xinyuan Qian,~\IEEEmembership{Senior Member,~IEEE}, Rohan Kumar Das,~\IEEEmembership{Senior Member,~IEEE}, \\
Xiaoxue Gao, Jiadong Wang, Haizhou Li,~\IEEEmembership{Fellow,~IEEE} 
}
\thanks{Ruijie Tao and Jiadong Wang are with the Department of Electrical and Computer Engineering, National University of Singapore, Singapore (e-mail: ruijie.tao@u.nus.edu and jiadong.wang@u.nus.edu)}
\thanks{Xinyuan Qian is with the Department of Computer Science and Technology, University of Science and Technology Beijing, Beijing, 100083, China. (email: qianxy@ustb.edu.cn)}
\thanks{Rohan Kumar Das is with Fortemedia, Singapore 138637 (e-mail: ecerohan@gmail.com)}
\thanks{Xiaoxue Gao is with Institute for Infocomm Research, A*STAR, Singapore. (email: Gao\textunderscore Xiaoxue@i2r.a-star.edu.sg)}
\thanks{Haizhou Li is with the Guangdong Provincial Key Laboratory of Big Data Computing, the Chinese University of Hong Kong (Shenzhen), 518172 China; also with  Shenzhen Research Institute of Big data, Shenzhen, 518172 China; also with the Department of Electrical and Computer Engineering, National University of Singapore, 119077, Singapore and also with the University of Bremen, 28359 Germany. (e-mail: haizhouli@cuhk.edu.cn)}
}

\begin{document}
\newcommand{\qian}[1]{{\color{red}#1}}

\newcommand{\jd}[1]{{\color{blue}#1}}
\include{acro}

\title{Enhancing Real-World Active Speaker Detection \\ with Multi-Modal Extraction Pre-Training}


\maketitle

\begin{abstract}
Audio-visual active speaker detection (AV-ASD) aims to identify which visible face is speaking in a scene with one or more persons. Most existing AV-ASD methods prioritize capturing speech-lip correspondence. However, there is a noticeable gap in addressing the challenges from real-world AV-ASD scenarios. Due to the presence of low-quality noisy videos in such cases, AV-ASD systems without a selective listening ability are short of effectively filtering out disruptive voice components from mixed audio inputs. In this paper, we propose a Multi-modal Speaker Extraction-to-Detection framework named `MuSED', which is pre-trained with audio-visual target speaker extraction to learn the denoising ability, then it is fine-tuned with the AV-ASD task. Meanwhile, to better capture the multi-modal information and deal with real-world problems such as missing modality, MuSED is modelled on the time domain directly and integrates the multi-modal plus-and-minus augmentation strategy. Our experiments demonstrate that MuSED substantially outperforms the state-of-the-art AV-ASD methods and achieves 95.6\% mAP on the AVA-ActiveSpeaker dataset, 98.3\% AP on the ASW dataset, and 97.9\% F1 on the Columbia AV-ASD dataset, respectively. We will publicly release the code in due course.

\end{abstract}

\begin{IEEEkeywords}
Audio-visual active speaker detection, audio-visual target speaker extraction, pre-training, self-supervised learning
\end{IEEEkeywords}

\section{Introduction}

\IEEEPARstart{I}n social interactions, humans rely on acoustic and visual inputs to understand the surrounding environments~\cite{865479, ephrat2018looking, 9858007}. While in complex multi-person scenarios, the first step is to know the speaking circumstances for each visible face clip at the visual frame-level~\cite{roth2020ava} (i.e., whether speaking or not). This task is referred to as audio-visual active speaker detection (AV-ASD), which serves an essential frontend for speech-related downstream tasks, such as audio-visual speaker recognition~\cite{tao2023self}, diarization~\cite{patrona2016visual, xu2022ava}, tracking~\cite{qian2022audio} and speech recognition~\cite{ afouras2018deep}. 

\begin{figure}[!ht]
    \centering
    \includegraphics[width=\linewidth]{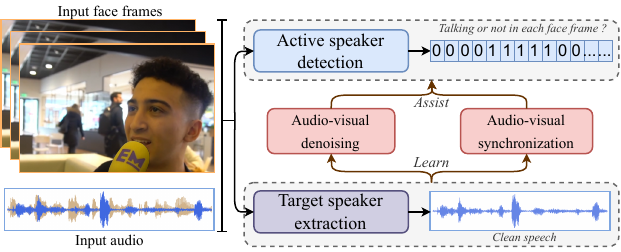}
    \caption{Audio-visual target speaker extraction can guide the neural network to learn the audio-visual denoising ability and synchronization knowledge, which can further assist the audio-visual active speaker detection task, especially in a challenging acoustic environment. In this example, the background is noisy, and the lip is occluded.}
    \vspace{-4mm}
    \label{fig:Intro_cover}
\end{figure}

In general, both intra-modality and inter-modality learning are significant for AV-ASD~\cite{roth2020ava}: \textit{Single-modality learning} is to detect if the voice heard belongs to human speech, and whether the lips of the interested person are moving. This step can exclude some non-talking segments. However, if both audio and visual modalities are active, \textit{Multi-modal learning} becomes necessary to detect the relationship between heard speech and lip movements. More specifically, human speech conveys phonemic information, and lip movements represent the viseme signal (the basic speech unit of the lip language). By matching them, the AV-ASD system can identify the talking person accurately.

Existing studies have shown that AV-ASD systems can perform well for high-quality videos with clean audio and visual inputs~\cite{min2022learning}, but real-world videos usually suffer from the noisy environment issue and the missing modality problem~\cite{afouras2019my, ma2021smil}. As shown in the left panel of Fig.~\ref{fig:Intro_cover}, the speech of interest may usually be accompanied by background noise, and the lip area might occasionally be obscured by hands or other objects during a conversation. In such scenarios, the AV-ASD model must isolate the acoustic signals of interest from the mixed audio under these challenging conditions, thereby relieving difficulties for multi-modal synchronization learning~\cite{voo2022delving, wang2024restoring}. However, existing AV-ASD researches predominantly focus on network architecture instead of the specific solutions for these real-world issues. We assert that these problems are crucial as they lead to a significant performance bottleneck for modern AV-ASD systems.

An efficient solution to provide the model with specific ability is to perform self-supervised learning (SSL), which pre-trains the system on large-scale unlabeled datasets followed by the task-specific fine-tuning stage~\cite{chen2020simple}. This method can leverage unlimited data and guide the model to learn the high-quality feature representation. However, to the best of our knowledge, there is no systematical AV-ASD-related pre-training solution: most of the speech-related SSL frameworks are conducted on one modality only, such as HuBERT~\cite{hsu2021hubert}, data2vec~\cite{baevski2022data2vec} and WavLM~\cite{chen2022wavlm} (for speech), and lip reading~\cite{afouras2018deep} (for vision). The most related multi-modal SSL framework is AV-HuBERT~\cite{shi2022avhubert}, which is learnt for speech recognition with high-level semantic information. However, AV-ASD is a fine-grained detection task focused more on the low-level temporal representation, i.g., predicting the instant talking status within every 40ms interval~\cite{roth2020ava}. In this paper, different from all previous frequency-domain solutions~\cite{tao2021someone, zhang2021unicon, min2022learning}, we designed the first time-domain AV-ASD network named `Multi-modal Speaker Extraction-to-Detection' (MuSED). It considers the audio-visual target speaker extraction (AV-TSE)~\cite{wu2019time} as the pre-training task to boost the framework for handling diverse real-world AV-ASD scenarios.

As shown in the right panel of Fig.~\ref{fig:Intro_cover}, the AV-TSE task aims to extract the specific person's voice from the mixture of speech based on the given speaker's lip movements~\cite{wu2019time, pan2020muse, pan2022usev}. As revealed in neuroscience studies~\cite{mesgarani2012selective}, humans have the ability to selectively listen to the voice of a specific speaker (refer to as target speaker) in a multi-talker environment, known as the `cocktail party effect'~\cite{bronkhorst2000cocktail}. This selective listening capability is associated with audio-visual denoising and speech-lip synchronization, which can assist the AV-ASD tasks. Therefore, AV-TSE and AV-ASD are highly related and share the common multi-modal representation information. However, combining these two tasks is difficult since they have different pipelines, framework structures and learning targets. Based on that, in our proposed MuSED framework, we design a unified structure that can be pre-trained with the AV-TSE task and fine-tuned with the AV-ASD task. Moreover, pre-training with AV-TSE can leverage the large-scale unlabelled talking videos since the mixed audio can be easily simulated by overlaying two speech utterances~\cite{ephrat2018looking}.

Additionally, we designed the AV-ASD augmentation strategy named the plus-and-minus strategy to further deal with the low-quality inputs during AV-ASD fine-tuning: The plus strategy adds the general natural sounds to the original audio input to boost the AV-ASD system. On the other hand, inspired by the masked auto-encoders (predict the masked image in SSL computer vision ~\cite{he2022masked}), the minus strategy applies masking and cropping operations during AV-ASD training. It randomly drops some components of the audio or visual inputs, which can force the network to reason for the missing part.

The contributions of our paper can be summarised as follows:

\begin{itemize}
\item We propose a new AV-ASD framework named Multi-modal Speaker Extraction-to-Detection (MuSED), which leverages the novel audio-visual pre-training technique for the AV-ASD task. MuSED applies self-supervised audio-visual target speaker extraction for pre-training to learn the multi-modal selective listening capability. After that, MuSED is fine-tuned for AV-ASD to detect the active speaker.

\item We design a unified model architecture in MuSED to match the pipeline of both AV-TSE and AV-ASD tasks. As the first time-domain AV-ASD framework, MuSED achieves comparable performance to the previous frequency-domain approaches. Meanwhile, the plus-and-minus strategy is applied to handle the noise and modality missing problems. 

\item We conduct extensive experiments to showcase the effectiveness of our proposal over the other state-of-the-art methods on three public datasets. Further, we ablate the key components of MuSED to show their contributions.
\end{itemize}

\section{Related work}

\subsection{Audio-visual active speaker detection}
AV-ASD aims to judge the talking status of each person based on multi-modal information~\cite{roth2020ava}. If multiple faces are on the screen, the AV-ASD system can handle each face individually with the face detector frontend. The earlier AV-ASD studies used to divide the videos into very short clips and predict the talking status of the interested person based on short-term temporal information~\cite{chung2016out, chung2019naver, alcazar2020active}. However, this setting ignores that long-term temporal information~\cite{shvets2019leveraging} is a key component of AV-ASD. Our previous TalkNet~\cite{tao2021someone} framework aims to capture the temporal dynamics of audio and visual streams as well as their interactions. This framework enables an accurate prediction of AV-ASD results. After that, \cite{datta2022asd} further investigates the transformer employment to TalkNet for better exploration of the temporal information and audio-visual correlation. While \cite{min2022learning} uses graph neural network (GNN) and treats AV-ASD as a binary node classification task to model the long-term spatial and temporal relationships among the cropped faces. Similarly, \cite{alcazar2022end} applies GNN to compose an end-to-end AV-ASD network that jointly learns the multi-modal feature and contextual predictions.

\begin{figure}[!b]
    \centering
    \includegraphics[width=\linewidth]{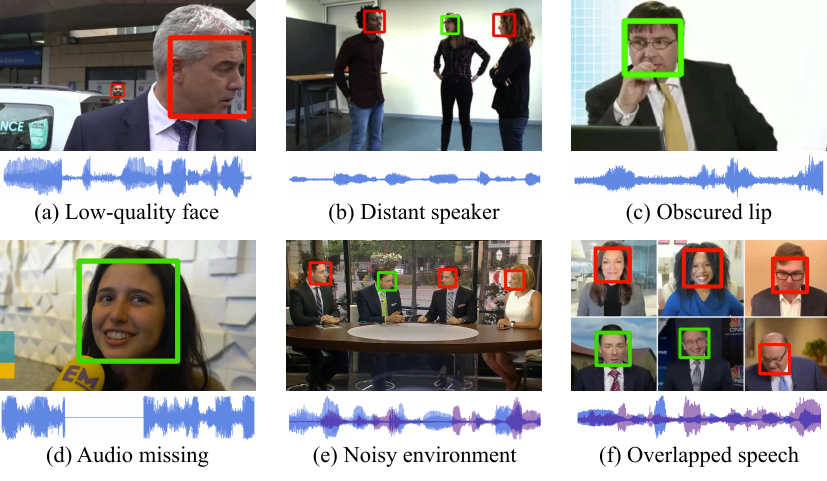}
    \caption{The illustration of the challenges of AV-ASD for the videos in the wild. The green and red boxes indicate the talking and non-talking speakers, respectively. (a) The video contains the side face and small face. (b) The speech is low-volume. (c) The lip is covered. (d) The audio segments are missed due to a microphone failure issue. (e) Noisy environment. (f) Multiply people talk concurrently with overlapped speech.}
    \label{fig:Related_cover}
\end{figure}

In summary, all previous AV-ASD studies are modelled on the frequency domain by utilizing analyzers such as MFCC~\cite{jiang23c_interspeech} or spectrogram~\cite{zhang2021unicon}. However, AV-ASD is a fine-grained waveform-level task to learn multi-modal correlation. The time-domain solution is also a reasonable choice with the advantage of capturing the temporal dynamics by modeling waveform directly. On the other hand, the primary challenges of AV-ASD occur from difficult realistic situations instead of high-quality videos, such as the noisy environments, ambiguous facial orientations, and complicated scenes~\cite{afouras2019my, voo2022delving, wang2024restoring} that are listed in Fig.~\ref{fig:Related_cover}. Based on that, our paper focuses on obtaining high-quality audio-visual embeddings from the defective videos and limited resources to deal with such challenging AV-ASD scenarios.

\vspace{-3mm}

\subsection{Audio-visual pre-training}
In speech processing and computer vision research, encoders usually apply specific CNN, RNN or Transformer structures to facilitate various task-specific applications~\cite{he2016deep, vaswani2017attention}. However, tasks from the same inputs share common aspects. To this end, substantial efforts have been paid to develop pre-trained models on the large-scale unlabeled corpus to learn universal speech~\cite{hsu2021hubert, chen2022wavlm} or image representations~\cite{he2022masked, dosovitskiy2020image}, which can eventually contribute to different downstream tasks by fine-tuning. Benefiting from abundant resources, these methods can understand the modality with more general and robust embeddings.

Existing pre-training studies usually focus on the single modality to learn the general feature presentation, such as text~\cite{bert}, speech~\cite{chen2022wavlm} and image~\cite{chen2020simple}. For audio-visual pre-training, \cite{afouras2018deep} leverages speech-lip synchronization to recognize speech content. AV-HuBERT~\cite{shi2022avhubert} captures cross-modal correlations by predicting the masked audio-visual inputs. The learnt audio-visual universal representations can benefit downstream tasks such as lip-reading and speech recognition. These pre-training methods usually concentrate on high-level semantic information. However, AV-ASD requires more for the low-level temporal speech-lip correlation~\cite{tao2021someone}. Due to that, pre-training with AV-TSE can be a suitable choice for AV-ASD task.

\vspace{-3mm}

\subsection{Audio-visual target speaker extraction}
As we mentioned, the challenge of AV-ASD is to handle the complex environment, where the model has to focus on the interested person's possibly existing speech. Motivated by that, we propose the extraction-based pre-train idea from the `cocktail party effect'~\cite{bronkhorst2000cocktail}: when the auditory environment comprises a complex mixture of several voices, AV-TSE treats the person of interest as the target speaker and performs selective listening to obtain his or her clean speech~\cite{wang2018voicefilter, xu2019time}. Here, the lip movements of the target speaker can provide valuable guidance for extraction through speech-lip correlation~\cite{afouras2018conversation}. 

Many algorithms, especially the time-domain solutions, have proved the feasibility and robustness of this multi-modal extraction idea, such as MuSE~\cite{pan2020muse}, USEV~\cite{pan2022usev} and TDSE~\cite{wu2019time}. These methods use a visual encoder to generate lip embedding sequences for the target speaker, which are then temporally aligned with speech frames for successful extraction. As the pre-training task, AV-TSE can leverage unlimited audio-visual samples in a self-supervised learning manner.

\section{Multi-modal Speaker Extraction-to-Detection framework (MuSED)}

\begin{figure}[!htb]
    \centering
    \includegraphics[width=\linewidth]{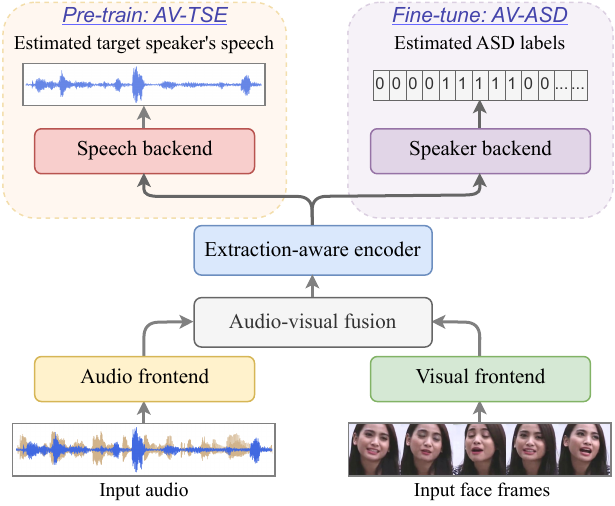}
    \caption{Overall illustration of our proposed MuSED framework. The left-upper branch: pre-training stage for audio-visual target speaker extraction (AV-TSE). The right-upper branch: fine-tuning stage for audio-visual active speaker detection (AV-ASD).}
    \label{fig:Method_overall}
    \vspace{-3mm}
\end{figure}

\subsection{Overall and task formula}

Our study proposes a time-domain AV-ASD framework named MuSED to predict the talking status of each face track. As shown in Fig.~\ref{fig:Method_overall}, MuSED is a unified framework that can match the format of the AV-ASD task (the left-upper panel) and AV-TSE task (the right-upper panel). Our MuSED is first pre-trained for AV-TSE with a self-supervised learning format. This stage guides the system in learning audio-visual synchronization and obtaining the ability to extract clean speech. Then, MuSED is fine-tuned for the AV-ASD task with the plus-and-minus data augmentation strategy to predict the speaking status accurately.

Let us define these two tasks first: For one video segment, considering that $v$ represents the continuous visual face frames from one interested speaker and $x$ is the heard audio simultaneously. $x$ can be a mixture of sounds that contains the target speaker's voice $s$ and the source of other sounds $n$. 

\begin{equation}
    x = s + \sum_{i=1}^N{n_i}
    \label{e1}
\end{equation}

$N$ is the number of other sound sources that $\geq 0 $. AV-ASD task aims to predict the target speaker's talking status in each visual frame with a binary score sequence $\hat{P}$, the length of $\hat{P}$ equals the number of visual frames. AV-TSE task aims to estimate the target person's clean speech $\hat{s}$ to approximate $s$.

\subsection{Time-domain audio-visual active speaker detection}
\begin{figure*}[!tb]
    \centering
    \includegraphics[width=\linewidth]{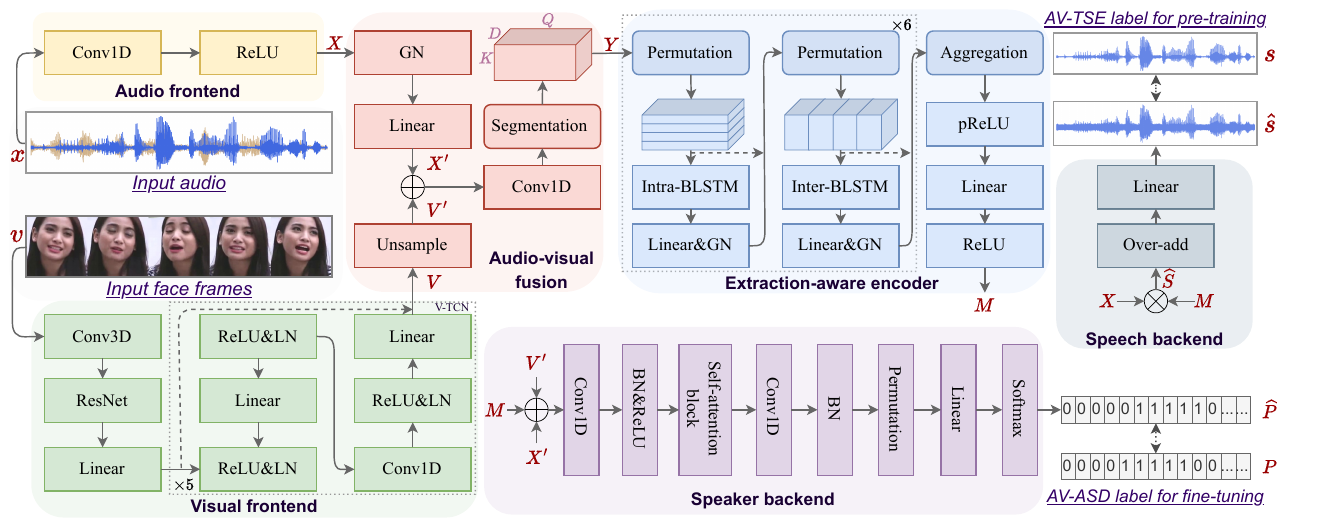}
    \caption{Detailed illustration of our proposed MuSED framework. It is first pre-trained with AV-TSE, then fine-tuned with AV-ASD task. MuSED contains the audio frontend, visual frontend, audio-visual fusion block, extraction-aware encoder, speech backend (used in AV-TSE pre-training only) and speaker backend (used in AV-ASD fine-tuning only). $\oplus$ and $\otimes$ represents concatenation and matmul product, respectively.}
    \label{fig:Method_details}
\end{figure*}

Our proposed time-domain AV-ASD design is motivated by the success of time-domain speech separation system USEV~\cite{pan2022usev, luo2020dual}. As shown in Fig~\ref{fig:Method_details}, for AV-ASD, MuSED contains the audio frontend, visual frontend, audio-visual fusion block, extraction-aware encoder and speaker backend.

\subsubsection{Audio frontend} For the audio modality, the input audio $x$ is an 1D waveform signal with the length equal to $T$. With a 1D-CNN and a rectified linear activation (ReLU), audio frontend can processes this signal and outputs a 2D embedding sequence $X$. The length of $X$ is equal to the amount of audio frames. It can intelligently translate the 1D temporal signal to the 2D representation with the learnable model structure.

\subsubsection{Visual frontend} For the visual modality, the inputs are the facial frames $v$ of the target speaker, where the specific viseme information implicitly appears in different mouth shape. Our visual frontend is a convolutional neural networks (CNN) based system. Utilizing a 3D-CNN layer, a ResNet18 block and a linear layer~\cite{afouras2018deep}, the local representation of each image can be comprehended. Subsequently, a video temporal convolutional block (V-TCN) is employed to integrate the temporal dynamics of lip movements and output the 2D visual embedding sequence $V$, the length of $V$ is equal to the number of visual frames. V-TCN has five residual-connected blocks, each block contains ReLU, layer normalization (LN), convolutional layer and linear layers~\cite{wu2019time}.

\subsubsection{Audio-visual fusion} The audio embedding $X$ is fed into a group normalization (GN) and a linear layer to adjust the feature dimension. This output embedding is denoted as $X^{\prime}$. Then the visual embedding $V$ is up-sampled to ${V^{\prime}}$ to align with the audio embedding $X^{\prime}$. $V^{\prime}$ and $X^{\prime}$ are concatenated along the time dimension and fed into a 1D-CNN layer. To address the variable length issue, the combined embedding is split into several temporal overlapping chunks~\cite{luo2020dual} with 50\% overlap rate. These chunks are concatenated together to form the 3D embedding $Y \in R ^ {D \times K \times Q}$, where $D$, $K$ and $Q$ represent the feature dimension, the length of each chunk and the number of chunks, respectively.

\subsubsection{Extraction-aware encoder} Six audio-visual extraction blocks with the same architecture are serially connected in the extraction-aware encoder to extract the high-quality feature of the target speaker's potential speech based on $Y$. In each block, the permutation operation is conducted first, and a Bidirectional Long Short-Term Memory (BLSTM) layer is performed within each chunk to integrate the local temporal knowledge. After a linear layer with GN, we re-order the feature with the permutation step and apply another BLSTM layer to integrate the global temporal information among all chunks. These extraction blocks are residual-connected for stable training. At last, this output 3D tensor is fed into the aggregation block, a parametric rectified linear activation (pReLU), a linear layer and a ReLU to output the 2D embedding $M$. $M$ is the multi-modal feature containing the speech signal representation that can correlate with the given lip movements.

\subsubsection{Speaker backend} An attention-based speaker backend is applied to achieve AV-ASD by modelling the audio-visual temporal dynamics and interactions. To better integrate the information from audio and visual modality, the audio feature $X^{\prime}$ and the visual feature $V^{\prime}$ are concatenated with the multi-modal feature $M$ together along the temporal dimension. This new feature is fed into the 1D convolutional block with BN and ReLU to adjust the temporal sampling rate, followed by the self-attention block to learn the global multi-modal temporal correlation~\cite{tao2021someone}. After that, another 1D convolutional block with BN and permutation operation is used to match the length of the output feature to the length of the AV-ASD labels. Finally, a fully connected layer with a softmax operation is used to generate the predicted AV-ASD label sequence. With this backend, MuSED can distinguish the talking and non-talking visual frames. 

\subsubsection{Loss function for AV-ASD} We treat AV-ASD as a frame-level speaking activity classification task, the training loss is the cross-entropy loss between the predicted label sequence $\hat{P}$ and the ground-truth label sequence $P$:
\begin{equation}
    \mathcal{L}_{AV-ASD} = - \frac{1}{{T}} \sum_{t=1}^{{T}} (\hat{p_t} \cdot \mathrm{log}\; p_t + (1-\hat{p_t}) \cdot \mathrm{log}\; (1-p_t))
    \label{formula:loss}
\end{equation}
where $\hat{p_t}$ and $p_t$ represent the predicted and ground-truth AV-ASD labels for the $t^{{\text {th}}}$ visual frame, respectively. Here, $t \in [1,T]$ and $T$ refers to the total number of visual frames.

\subsection{Pre-training with audio-visual target speaker extraction}
Pre-training MuSED with the AV-TSE can benefit the AV-ASD system in handling the challenging acoustic environment. To match the AV-ASD format, the pre-train stage utilizes the same audio frontend, visual frontend, audio-visual fusion block and extraction-aware encoder. The difference is that the speaker backend is removed and replaced by a speech backend to extract the possibly existing clean speech from the target speaker.

\subsubsection{Self-supervised learning data generation} Different from the AV-ASD task where the input videos have already been provided, we design a self-supervised learning pipeline to generate samples for AV-TSE pre-training: Consider one corpus with a large number of audio-visual synchronized videos. Each video contains only one visible talking face. Its audio and visual parts can be viewed as the target speech and the target speaker's face frames, respectively. Then, we randomly select one other video's audio as the interference voice and add it to the target speech under the specific signal-to-noise ratio to obtain the different kinds of mixed audio. Based on this pipeline, we can generate unlimited samples, where each sample contains the mixed audio $x$, the target speaker's lip movements $v$ and the target speaker's clean speech $s$ for AV-TSE pre-training.

\subsubsection{Speech backend} As shown in Fig.~\ref{fig:Method_details}, during AV-TSE per-training, the speech backend is applied to recover the time-domain speech signal from the 2D feature. In this backend, the learnt multi-modal feature $M$ is viewed as the mask of the target speaker's clean speech. It is multiplied with the audio feature $X$ to obtain the 2D speech feature $\hat{S}$, followed by an overlap-and-add (Over-add) operation and a linear layer. This speech backend can transform the 2D feature into the extracted 1D temporal speech signal $\hat{s}$. The talking segments have high energy, and the non-talking segments have low energy. Due to that, AV-TSE pre-training can assist the AV-ASD objective.

\subsubsection{Loss function for AV-TSE} In AV-TSE pre-training, to reduce the reconstruction error between the estimated speech $\hat{s}$ and the ground-truth speech $s$, the scale-invariant signal-to-distortion ratio (SI-SDR) loss~\cite{le2019sdr} is used:

\begin{equation}
        \mathcal{L}_{AV-TSE} = - 10 \log_{10} ( \frac{||\frac{<\hat{s},s>s}{||s||^2}||^2}{||\hat{s} - \frac{<\hat{s},s>s}{||s||^2}||^2})
\end{equation}

\subsubsection{Plus-and-minus augmentation strategy} 

The robustness of the AV-ASD system can be easily affected by noisy acoustic scenarios and the missing modality problem. We propose a systematic data augmentation strategy for AV-ASD named a plus-and-minus strategy. It is noted that the person's talking activity labels (i.e., speaking or not) will not be affected by our augmentation process.

\begin{figure}[!tb]
    \centering
    \includegraphics[width=\linewidth]{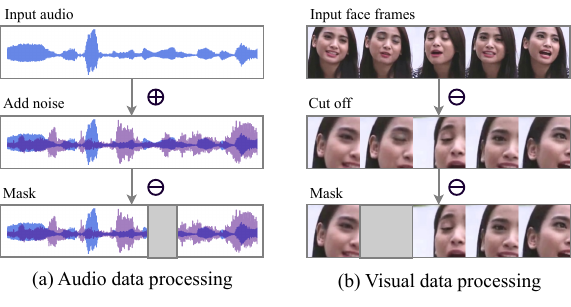}
    \caption{Plus-and-minus data augmentation strategy for AV-ASD training. For audio and visual modality, several operations are applied to imitate the noisy environment and missing modality problems to boost the AV-ASD performance under the challenge videos. $\oplus$ and $\ominus$ represents `plus' and `minus' operation, respectively.}
    \label{fig:Method_pm}
    \vspace{-4mm}
\end{figure}

As shown in Fig.~\ref{fig:Method_pm}, during AV-ASD training, the plus strategy adds one noisy waveform to the original waveform. This noisy waveform can come from any audio waveform in the AV-ASD training dataset (to introduce the in-domain noise), or the reverberation signals collected from the different rooms or spaces (to simulate different location scenes), or the ambient sounds, such as nature noises (the sounds from thunder, rain, ball, etc.), background music (playing the instrument or singing) and babble (a group of people talking at the same time). A diverse real-world acoustic environment can be simulated by adding this noisy signal. Then, we do the minus strategy by randomly masking some short audio segments to force the system to reason the missing parts and handle the missing audio condition.

On the other hand, a two-step minus strategy is applied for the visual inputs: The cut-off step generates a square box with a random side length and crops the same region for all the visual frames in each video to mimic the partial face inputs. Next, we mask several continuous visual frames to simulate the face detection failures. With these two simple steps, the AV-ASD system must predict and understand the missed information based on the audio and visual context during AV-ASD training. It can reduce the impact of the missing modality problem.

\section{Experimental Setup}
\subsection{Dataset}

Three popular AV-ASD datasets are used in our experiments: the AVA-ActiveSpeaker dataset~\cite{roth2020ava}, the Active Speaker in the Wild dataset~\cite{kim2021look} and the Columbia Active Speaker dataset~\cite{chakravarty2016cross}. Meanwhile, AV-TSE pre-training is conducted on our generated VoxCeleb2-Mix dataset.

\subsubsection{VoxCeleb2-Mix}
Any datasets with the single speaker's clean speech can be used for AV-TSE training. VoxCeleb2~\cite{chung2018voxceleb2} is the audio-visual speaker recognition dataset collected from the YouTube interviews. There are over one million videos from 5,994 speakers in the dataset. Each video clip lasts more than 4 seconds and contains only one visible talking face. The heard speech and the lip movements are relatively synchronized. Based on that, we generate the AV-TSE pre-training dataset named VoxCeleb2-Mix and set the signal-to-noise ratio (SNR) of the generated mixture of speech between -10 and 10. There are 200,000 samples selected in VoxCeleb2 to generate VoxCeleb2-Mix. The interference signal can come from VoxCeleb2 itself, the MUSAN dataset (the dataset with natural noise), or both of them (related ablation experiments can be found in Table ~\ref{tab:Results_pre_train}). As shown in Table ~\ref{tab:Experiments_dataset}, we fine-tuned MuSED on three AV-ASD datasets for evaluation after pre-training.

\begin{table}
    \caption{A summary of the datasets used for AV-ASD training and evaluation. The duration of these datasets is also provided. AV-TSE pre-training is on the VoxCeleb2-Mix dataset.}
    \centering
    
    \begin{tabular}{p{2cm}<{\centering}p{1cm}<{\centering}|p{2cm}<{\centering}p{1cm}<{\centering}}
        \hline
        \multicolumn{2}{c}{Training Set} & \multicolumn{2}{c}{Evaluation Set} \\
        \hline
        Dataset & Hours & Dataset & Hours \\
        \hline
        AVA       & 29.02  & AVA  & 8.23 \\
        ASW       & 13.37  & ASW  & 7.91\\
        TalkSet   & 151.65 & CAS  & 1.45 \\
         \hline
    \end{tabular}    
    \label{tab:Experiments_dataset}
    \vspace{-2mm}
\end{table}

\subsubsection{AVA-ActiveSpeaker (AVA)}
The AVA-ActiveSpeaker dataset~\cite{roth2020ava} contains videos from movies and has a total of 29,723, 8,015, and 21,361 face tracks in the training, validation, and test sets, respectively\footnote{\url{https://research.google.com/ava/download.html##ava_active_speaker_download}}. These tracks range in length from 1 to 10 seconds. Since the evaluation for the test set is unavailable now, following all previous works, we use the official evaluation tool to report the mean average precision (mAP) on its validation set. The AVA dataset is challenging due to its diverse languages.

\subsubsection{Active Speaker in the Wild (ASW)}
The ASW dataset~\cite{kim2021look} includes videos from YouTube interviews. The dataset is divided into training, validation, and test sets, with 4,676, 3,483, and 3,392 face tracks\footnote{\url{https://github.com/clovaai/lookwhostalking}}. The minimum duration of a track in the dataset is 2.1 seconds, while the maximum is 311.0 seconds. Specifically, we split all long tracks into short utterances of 20 seconds or less due to memory issues. Following the existing works, we report average precision (AP), the area under the receiver operating characteristic (AUROC), and the equal error rate (EER) on its test set.

\subsubsection{Columbia Active Speaker (CAS)}
The CAS dataset~\cite{chakravarty2016cross} consists of an 87-minute video of a panel discussion featuring 5 speakers who take turns speaking, with 2 to 3 speakers visible at any given time\footnote{\url{http://www.jaychakravarty.com/active-speaker-detection/}}. Following the common protocol for this benchmark, we evaluate the model using the F1 score. However, due to the limited size of the CAS dataset, it is usually used as a test set. Therefore, we generate TalkSet dataset~\cite{tao2021someone} as the training set. This dataset covers all five valid AV-ASD scenarios. In total, we generated 150,000 videos with 151.65 hours from VoxCeleb2 and LRS3 datasets, where 71.45 hours are talking and 80.20 hours are non-talking. We randomly pick 135,000 training samples and 15,000 validation samples.

\subsection{Implementation Details}
We used Adam as the optimizer in all the experiments. For AV-TSE pre-training, all samples are cut into 4 seconds, and the entire structure is trained from scratch; for AV-ASD fine-tuning, data with variable lengths are used for training. In all stages, each face frame contains one single face of the target speaker. All frames are converted into grey images and reshaped into $112\times112$ as the visual inputs.

For the plus strategy, the reverberation data comes from the RIRS dataset~\cite{RIRS}, and the ambient noise data comes from the MUSAN dataset~\cite{MUSAN}. For the minus strategy, the side length of the cut square box is randomly selected between 45 and 112 pixels. Then, for both the audio and visual modality, we view the videos as several segments of the same duration (this length is randomly selected between 0.2 and 0.8 seconds). Then, we mask 25\% (for audio) and 10\% (for visual) of these segments with zeros during AV-ASD training, respectively. More details can be found in our soon-to-be-open source code.

\section{Results and Analysis}

\subsection{Results of MuSED on the AVA Dataset} 
Table~\ref{tab:Results_AVA} reports the performance of MuSED on the AVA dataset and compares it to the previous works. MuSED achieves a 95.6\% mAP on AVA and illustrates an improvement of and 0.7\% over the state-of-the-art method  (SPELL+~\cite{min2022learning}). Moreover, the existing works usually apply a two-stage~\cite{min2022learning} or even a three-stage~\cite{kopuklu2021design} system to study the auxiliary information (relationship between the speakers who appear at the same time). In contrast, our MuSED is a single stage framework with a relative light-weight model structure that predicts the target speaker's talking status directly. Therefore, there lies scope to improve performance of proposed MuSED with these multi-stage strategies. However, as a frontend for other downstream audio-visual tasks, a simple AV-ASD pipeline might be a more suitable solution. 

\begin{table}[!h]
  \caption{Comparison with the state-of-the-art methods on the AVA dataset in terms of mean average precision (mAP). The number of parameters in each method is also provided.}  
  \begin{tabular}{p{3.2cm}<{\centering}p{2cm}<{\centering}p{2cm}<{\centering}}
    \hline
    \textbf{Method} &  \textbf{mAP (\%)} & \textbf{Params (M)} \\
    \hline
    Zhang et al.~\cite{zhangmulti} & $84.0$& $22$ \\
    ASC~\cite{alcazar2020active} & $87.1$& $23.5$ \\
    Chung et al.~\cite{chung2019naver} & $87.8$& $13$ \\
    MAAS-TAN~\cite{leon2021maas} &$88.8$&  $22.5$ \\
    Pouthier et al.\cite{pouthier2021active} & $91.9$& $2.0$ \\
    UniCon~\cite{zhang2021unicon} & $92.0$& $23.8$ \\
    TalkNet~\cite{tao2021someone}  & $92.3$& $15.7$ \\
    ASDNet~\cite{kopuklu2021design}  & $93.5$& $51.3$ \\
    EASEE-50~\cite{alcazar2022end}  & $94.1$& $>74.7$ \\
    Light-ASD~\cite{liao2023light}  & $94.1$ &  $1.0$ \\
    SPELL~\cite{min2022learning}  & $94.2$& $22.5$ \\
    UniCon+~\cite{zhang2022unicon+} & $94.5$& $>23.8$ \\
    SPELL+~\cite{min2022learning}  & $94.9$& $>45.0$\\
    \textbf{MuSED (ours)}  & $\textbf{95.6}$& $16.1$ \\
    \hline
  \end{tabular}
  \label{tab:Results_AVA}
\end{table}

\subsection{Results of MuSED on the ASW Dataset} 

\begin{table}[!t]
  \caption{Comparison with the state-of-the-art methods on ASW dataset.}
  \begin{tabular}{p{2.4cm}<{\centering}p{1.4cm}<{\centering}p{1.9cm}<{\centering}p{1.4cm}<{\centering}}
    \hline
    \textbf{Method} & \textbf{AP (\%) $\uparrow$} & \textbf{AUROC (\%) $\uparrow$} & \textbf{EER (\%) $\downarrow$}\\
    \hline
    Roth et al.~\cite{roth2020ava, wuerkaixi2022rethinking} & $89.7$ & $-$ & $-$\\
    SynNet~\cite{chung2016out, wuerkaixi2022rethinking} & $92.4$ & $-$ & $-$\\
    ASW-SSL~\cite{kim2021look} & $90.5$ & $95.1$ & $9.8$\\
    Wang's~\cite{wang2022lip} & $93.4$ & $95.4$ & $10.4$\\
    ASW-BGRUs~\cite{kim2021look} & $96.6$ & $97.2$ & $6.2$\\
    TalkNet~\cite{tao2021someone} & $97.7$ & $98.6 $ & $5.1$\\
    \textbf{MuSED (ours)} & $\textbf{98.3}$ & $\textbf{99.0} $ & $\textbf{3.6}$\\
    \hline
  \end{tabular}
  \label{tab:Results_ASW}
\end{table}

Table~\ref{tab:Results_ASW} shows the performance of MuSED and compares it with other works on the ASW dataset. We observe that MuSED achieves the highest AP and AUROC with 98.3\% and 99.0\%, respectively. Meanwhile, it also obtains the lowest EER (3.6\%). For all metrics, MuSED outperforms the best existing system. Considering that all samples in ASW are in-the-wild videos, this result demonstrates practical application value of MuSED in addressing the real-world AV-ASD problem.

\subsection{Results of MuSED on the CAS Dataset} Table~\ref{tab:Results_CAS}  reports the results of MuSED on the CAS dataset. The F1 score is reported for each of the five speakers in the dataset. Noted that a small number of AV-ASD labels of the speaker `Bell' and `Boll' are incorrect: the label shows the speaker is talking, but only the lip of the speaker is moving. Table~\ref{tab:Results_CAS} reports the results of MuSED on our manually calibrated labels. \footnote{We upload the corrected ground-truth labels on \url{https://drive.google.com/drive/folders/1xDuNUQDX7BE5eMDsW46ADeXOiW7kZafR?usp=sharing}}. Among all approaches, our MuSED obtains the highest average F1 score of 97.4\%, which outperforms the best previous system with an absolute improvement of 1.2\%. These experiments demonstrate that our MuSED achieves state-of-the-art results on all three benchmark datasets.

\begin{table}[t!]
  \caption{Comparison with the state-of-the-art methods on the CAS dataset in terms of F1 scores (\%).}
  \begin{tabular}{p{2.3cm}<{\centering}p{0.6cm}<{\centering}p{0.6cm}<{\centering}p{0.6cm}<{\centering}p{0.6cm}<{\centering}p{0.6cm}<{\centering}|p{0.6cm}<{\centering}}
    \hline
    \multirow{2}{*}{\textbf{Method}} & \multicolumn{6}{c}{\textbf{Speaker Name}}\\
    {} & \textbf{Bell} & \textbf{Boll} & \textbf{Lieb} & \textbf{Long} & \textbf{Sick} & \textbf{Avg.}\\
    \hline
    Punarjay et al.~\cite{chakravarty2016cross} & $82.9$ & $65.8$ & $73.6$ & $86.9$ & $81.8$ & $78.2$~~~\\
    Zach et al.~\cite{shahid2019comparisons} & $89.2$ & $88.8$ & $85.8$ & $81.4$ & $86.0$ & $86.2$~~~\\
    RGB-DI~\cite{shahid2019comparisons} & $86.3$ & ${93.8}$ & $92.3$ & $76.1$ & $86.3$ & $87.0$~~~\\
    SyncNet~\cite{chung2016out} & $93.7$ & $83.4$ & $86.8$ & ${97.7}$ & $86.1$ & $89.5$~~~\\
    LWTNet~\cite{afouras2020self} & $92.6$ & $82.4$ & $88.7$ & $94.4$ & $95.9$ & $90.8$~~~\\
    RealVAD~\cite{beyan2020realvad} & $92.0$ & ${98.9}$ & $94.1$ & $89.1$ & $92.8$ & $93.4$~~~\\
    S-VVAD~\cite{shahid2021s} & $92.4$ & $97.2$ & $92.3$ & $95.5$ & $92.5$ & $94.0$~~~\\
    GSCMIA\cite{sharma2022audio}     & $97.3$ & $96.3$ & $89.4$ & $98.7$ & $98.7$ & $96.2$~~~\\
    TalkNet\cite{tao2021someone}     & $97.1$ & $90.0$ & ${99.1}$ & $96.6$ & $98.1$ & $96.2$~~~\\
    \textbf{MuSED (ours)}     & ${98.5}$ & $94.7$ & $96.3$ & ${99.3}$ & ${98.0}$ & $\textbf{97.4}$~~~\\    
    \hline
  \end{tabular}  
  \vspace{2mm}
  \label{tab:Results_CAS}
\end{table}

\subsection{Ablation study for the MuSED framework}
Then we conduct ablation study to investigate the efficiency of each algorithm that proposed in MuSED, which includes AV-TSE pre-training, time-domain architecture and plus-and-minus strategy. 

\begin{table}[ht]
\centering
  \caption{The study of AV-TSE pre-training. The upper part lists the previous AV-ASD works with different pre-train strategies; the bottom part studies the different type of interference signal during AV-TSE pre-training: `-N', `-S', '-SN' and '-SSN' represent the non-speech noise, the speech from one other speaker, the summation of them, and the summation of two speakers' speech plus the non-speech noise, respectively.}
  
  \label{tab:pretrain}
  \begin{tabular}{p{2cm}<{\centering}p{1.2cm}<{\centering}p{3.2cm}<{\centering}p{.8cm}<{\centering}}
    \hline
    \multirow{2}{*}{Method} & Pre-train & Pre-train & \multirow{2}{*}{mAP (\%)} \\
                            & modality & strategy &  \\
    \hline    
    Zhang et al.~\cite{zhangmulti} & V & Lip reading & $84.0$ \\
    ASC~\cite{alcazar2020active} & V & Image recognition & $87.1$ \\
    Chung et al.~\cite{chung2019naver} & AV & AV-synchronization & $87.8$ \\
    MAAS-TAN~\cite{leon2021maas} & V & Image recognition & $88.8$ \\
    ASDNet~\cite{kopuklu2021design} & V & Image \& action recognition & $93.5$  \\
    EASEE-50~\cite{alcazar2022end} & V & Image \& action recognition & $94.1$  \\
    SPELL~\cite{min2022learning}  & V & Image recognition & $94.2$  \\    
    \hline
    \multirow{4}{*}{MuSED (ours)} & \multirow{4}{*}{AV} & AV-TSE-N   & 93.5 \\
                                  & & AV-TSE-S   & 95.4 \\
                                  & & AV-TSE-SN  & 95.4 \\
                                  & & AV-TSE-SSN & \textbf{95.6} \\
    \hline                                                          
  \end{tabular}
  \label{tab:Results_pre_train}
\end{table}

\subsubsection{Efficiency of AV-TSE pre-training} First, we study whether AV-TSE is an efficient pre-training method for AV-ASD task. In the upper part of Table~\ref{tab:Results_pre_train}, we list the previous AV-ASD methods with different pre-train strategies for comparison. Most of them pre-trained the visual frontend with image or action recognition task. It can make the AV-ASD system better understand single modality but ignore the multi-modal correlation. Another way is to pre-train the system for AV synchronization~\cite{chung2019naver}. However, the information learned from synchronization is duplicated with the knowledge learnt in AV-ASD fine-tuning. Compared with them, our proposed AV-TSE pre-train mechanism guides the AV-ASD system to comprehend the multi-modal extraction ability and achieves the best performance with an mAP of 95.6\% by handling the challenging environments.

\subsubsection{Intensity of pre-training} We further study the significance of this extraction ability in AV-ASD. In the bottom part of Table~\ref{tab:Results_pre_train}, we report the results of MuSED with four different kinds of interference signals during Av-TSE pre-training: `-N', `-S', '-SN' and '-SSN' represent the interference signal is the non-speech noise, the speech from one other speaker, the summation of them, and the summation of two speakers' speech plus the non-speech noise, respectively. We notice that pre-training with heavy interference signals can benefit more for the MuSED system. That matches our motivation that AV-TSE pre-training provides the system with the denoising and selective listening ability. The stronger this capability system obtains, the better performance it can achieve to handle challenging conditions. 

\begin{table}[!htb]
    \centering
    \vspace{-4mm}
    \caption{The ablation study for the time-domain solution on the AVA dataset, TalkNet is viewed as the frequency-domain baseline for comparison. This table also studies the effect of the plus-and-minus strategy in MuSED.}
    \begin{tabular}{p{1.5cm}<{\centering}p{1cm}<{\centering}p{1.4cm}<{\centering}p{1.5cm}<{\centering}p{1cm}<{\centering}}
        \hline
        \multirow{2}{*}{Method} & \multirow{2}{*}{Pre-train} & Plus & Minus & \multirow{2}{*}{mAP (\%)}  \\
                                &            & strategy & strategy & \\
        \hline
        TalkNet~\cite{tao2021someone} & $\times$     & $\times$        & $\times$     & 89.4 \\
        \hline
        \multirow{6}{*}{MuSED (ours)} & $\times$     & $\times$        & $\times$     & 91.6 \\
                                & $\times$ & $\checkmark$        & $\checkmark$ & 93.1 \\
                                & $\checkmark$ & $\times$        & $\times$     & 93.1 \\
                                & $\checkmark$ & $\checkmark$    & $\times$     & 92.7 \\
                                & $\checkmark$ & $\times$        & $\checkmark$ & 94.6 \\
                                & $\checkmark$ & $\checkmark$    & $\checkmark$ & \textbf{95.6} \\                              
        \hline        
    \end{tabular}
    \label{tab:Results_pm}
\end{table}

\begin{figure}[!th]
    \centering
    \includegraphics[width=.95\linewidth]{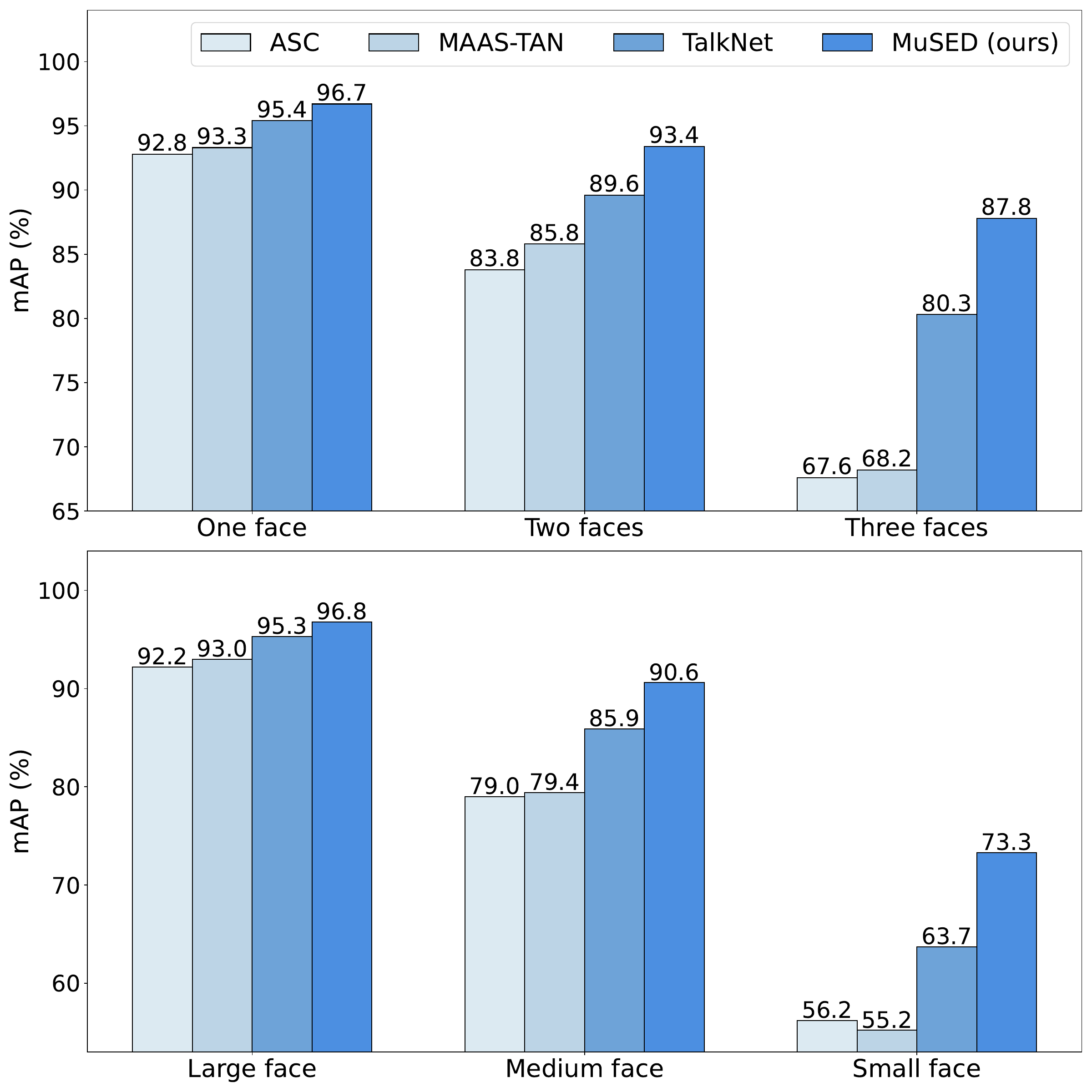}
    \caption{The mAP (\%) of MuSED and previous methods on the AVA dataset. The upper figure studies the results for different numbers of faces in the same visual frame; the bottom one studies the performance for the different face sizes.}
    \label{fig:Results_123sml}
\vspace{-4mm}
\end{figure}

\subsubsection{Time-domain vs frequency-domain} To verify that our time-domain AV-ASD solution is comparable to the previous frequency-domain method. In the first line of Table~\ref{tab:Results_pm}, we select TalkNet~\cite{tao2021someone} as the frequency-domain baseline for comparison on the AVA dataset since it has a similar model size with MuSED. Under a fair comparison without the per-train and data augmentation methods, MuSED achieves an 2.2\% mAP improvement compared to TalkNet with the same training setting, which proves that the time-domain solution is also competitive in AV-ASD task.

\subsubsection{Plus-and-minus strategy} Table~\ref{tab:Results_pm} also studies the effect of the plus-and-minus strategy. For MuSED with pre-training, we notice that the plus-and-minus strategy can obviously boost the system (93.1\% vs 95.6\% mAP). Compared with the plus strategy, the MuSED benefits more from the minus strategy. These results prove the efficiency of our strategy.

\begin{figure*}[!b]
    \centering
    \includegraphics[width=.95\linewidth]{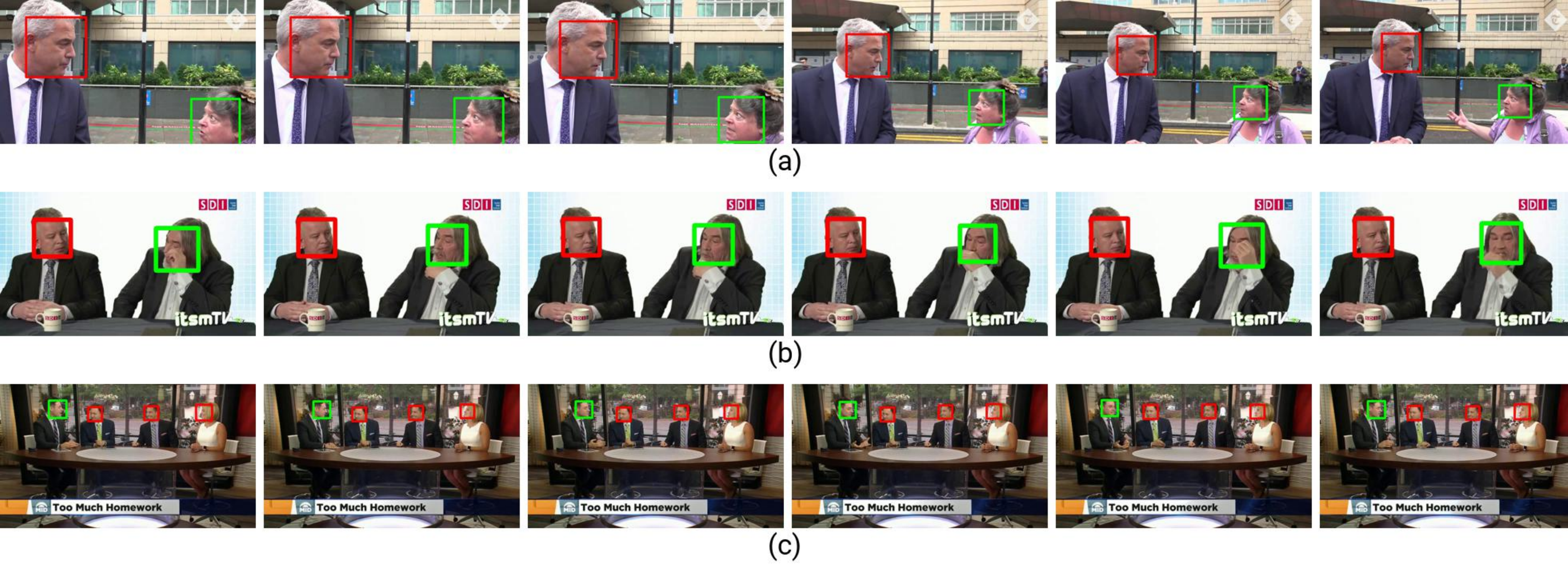}
    \caption{Examples of the MuSED on the challenge videos. These visual frames are arranged along timestamps. The green box denotes the active speaker. The red box denotes the inactive speaker. (a) Two speakers with side faces; (b) Two speakers take turns speaking, and the man’s lips are covered sometimes; (c) A meeting scenario with four small faces.}
    \label{fig:Results_visualization}
\end{figure*}

\vspace{-4mm}

\subsection{Quality study for the challenge scenarios}
In this section, we examine the robustness of MuSED on the AVA dataset under various test settings. We focus on the performance of MuSED for handling challenge conditions.

\subsubsection{Number of faces} Our MuSED analyzes individual face tracks independently when multiple people are on the screen. Here, the multi-people scenario usually indicates a more noisy environment. The upper part of Fig.~\ref{fig:Results_123sml} reports the results of MuSED under the different numbers of visible faces on the screen. Among all the approaches, MuSED always holds the best results for all the listed conditions. Meanwhile, compared with the simple conditions (1 or 2 faces on the screen), MuSED achieves more obvious improvements for the difficult situations (3 faces on the screen). These results demonstrate MuSED's capability to handle diverse and complex situations.

\subsubsection{Size of faces} The size of each original face track decides the difficulty of the AV-ASD task. Usually, it is more challenging to capture the lip movements of the smaller faces. As depicted in the bottom part of Fig.~\ref{fig:Results_123sml}, MuSED demonstrates the best performance across all face sizes in the evaluation set, including small faces (less than 64 pixels in width), medium faces (between 64 and 128 pixels in width), and large faces (more than 128 pixels in width). These results verify that MuSED can perform well for low-quality face tracks.

\subsubsection{Low-quality audio inputs} Table~\ref{tab:Results_low_a} reports the quality study for the low-quality acoustic conditions. During AV-ASD evaluation, we involve three kinds of additional voice: noise (sounds from nature), music or babble (whisper from multiple person),  into the original input waveform to simulate the different noisy environment. We notice that both AV-TSE pre-training and plus-and-minus augmentation strategy can boost performance, this improvement is more obvious for the very noisy scenarios with lower SNR.

\begin{table}[!t]
    \centering
    \caption{Robustness study for MuSED in the complex acoustic environment on the AVA dataset with mAP (\%), additional voice is involved with the different signal-to-noise ratio (SNR) during evaluation to simulate low-quality audio inputs. `PM' denotes plus-and-minus strategy during AV-ASD training.}
    \begin{tabular}{p{1.2cm}<{\centering}p{1cm}<{\centering}p{1cm}<{\centering}p{.6cm}<{\centering}p{.6cm}<{\centering}p{.6cm}<{\centering}p{.6cm}<{\centering}}
        \hline
        Voice  & Pre-train & PM & \multicolumn{4}{c}{SNR (dB)} \\
        type & strategy & strategy & 20 & 15 & 10 & 5  \\
        \hline
        \multirow{3}{*}{Noise} & $\checkmark$ & $\times$      & 92.98 & 92.78 & 92.39 & 91.59 \\
        & $\times$ & $\checkmark$      & 93.07 & 92.96 & 92.73 & 92.20 \\
        & $\checkmark$  & $\checkmark$ & \textbf{95.52} & \textbf{95.42} & \textbf{95.23} & \textbf{94.79} \\

        \hline
        \multirow{3}{*}{Music} & $\checkmark$    & $\times$  & 93.02 & 92.82 & 92.35 & 91.48 \\
        & $\times$    & $\checkmark$  & 93.06 & 92.95 & 92.62 & 91.86 \\
        & $\checkmark$ & $\checkmark$ & \textbf{95.53} & \textbf{95.44} & \textbf{95.22} & \textbf{94.75} \\

        \hline
        \multirow{3}{*}{Babble} & $\checkmark$    & $\times$  & 91.37 & 89.40 & 86.13 & 81.54 \\
        & $\times$    & $\checkmark$ & 92.54 & 91.66 & 89.39 & 85.51 \\
        & $\checkmark$ & $\checkmark$ & \textbf{95.18} & \textbf{94.51} & \textbf{92.92} & \textbf{89.98} \\

        \hline
    \end{tabular}
    \label{tab:Results_low_a}
\end{table}

\begin{table}[!htb]
    \centering
    \caption{Robustness study for MuSED evaluation with mAP (\%) when different proportions of visual frames are occluded by hand or object image. `pm' denotes plus-and-minus strategy.}
    \begin{tabular}{p{1.2cm}<{\centering}p{1cm}<{\centering}p{1cm}<{\centering}p{.6cm}<{\centering}p{.6cm}<{\centering}p{.6cm}<{\centering}p{.6cm}<{\centering}}
        \hline
        Occlusion  & Pre-train & PM & \multicolumn{4}{c}{\#\% of frames are occluded} \\
        type & strategy & strategy & 5 & 10 & 15 & 20  \\
        \hline
        \multirow{3}{*}{Hand} & $\checkmark$ & $\times$        & 92.67 & 92.32 & 91.99 & 91.52    \\
        & $\times$ & $\checkmark$        & 92.72 & 92.56 & 92.36 &   92.10  \\
        & $\checkmark$ & $\checkmark$    & \textbf{95.39} & \textbf{95.23} & \textbf{95.07} & \textbf{94.84}    \\
        
        \hline
        \multirow{3}{*}{Object} & $\checkmark$ & $\times$       & 92.30 & 91.80 & 91.21 & 90.34   \\
         & $\times$ & $\checkmark$       & 92.67 & 92.33 & 91.84 & 91.25   \\
         & $\checkmark$ & $\checkmark$  & \textbf{95.32} & \textbf{95.00} & \textbf{94.58} & \textbf{94.04}    \\

        \hline
    \end{tabular}
    \label{tab:Results_low_v}
\end{table}

\subsubsection{Low-quality visual inputs} 
\begin{figure}[t]
    \centering
    \includegraphics[width=.9\linewidth]{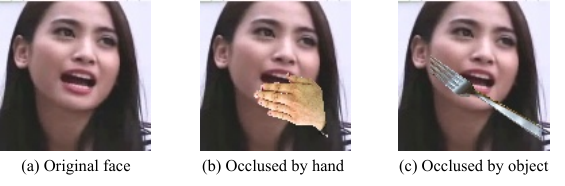}
    \caption{The example of the input frames in the study of low-quality visual inputs. The hand or object image is randomly chosen to cover the face image to simulate the occluded face.}
    \label{fig:Results_occlusion}
    \vspace{-5mm}
\end{figure}

Similarly, we conduct the quality study for the low-quality visual inputs~\cite{voo2022delving} in Table~\ref{tab:Results_low_v}. As shown in Fig.~\ref{fig:Results_occlusion} to simulate the occluded face, we randomly select one hand image from the Hand11K dataset~\cite{afifi201911k} or one object image from the COCO dataset~\cite{lin2014microsoft}, then overlap it to the original face frame during evaluation. We test the results of different percentages of the covered visual frames, where a higher percentage represents the visual information with lower quality. The results show that both pre-training and augmentation strategy can boost the system. These studies show that MuSED can address diverse conditions and excel in real-world video scenarios.

\subsection{Visualization}

At last, we collect some real-world challenging videos and provide the visualization results of our proposed MuSED in Fig.~\ref{fig:Results_visualization}. The green box represents the talking person, and the red box represents the non-talking person. These videos cover diverse real-world conditions such as the small face, side face, noisy environment, covered lip areas, and multiple speakers. MuSED can provide robust results with accurate and stable detection, while the previous methods can hardly deal with them.

\section{Conclusion}
In this work, we propose an audio-visual target speaker extraction-based pre-training mechanism. By obtaining selective listening ability from pre-training, MuSED, which is our time-domain audio-visual active speaker detection system, can handle real-world challenge videos. The experimental results show that the proposed MuSED outperforms the state-of-the-art methods on three public AV-ASD datasets. Our study provides the AV-ASD task with new inspiration and direction to explore the multi-modal pre-training technique and time-domain model structure. In future work, we plan to extend the AV-ASD module with the speaker identity information for the audio-visual speaker diarization system. It can recognize `who spoke when' for better social interaction. Meanwhile, we will further study the efficiency of our proposed multi-modal extraction-based pre-training method for other multi-modal tasks to check its generality, such as noise-robust audio-visual automatic speech recognition.

\bibliographystyle{IEEEtran}
\bibliography{ref}

\end{document}